\documentclass[twocolumn,showpacs,preprintnumbers,amsmath,amssymb,prb]{revtex4}
\usepackage{graphicx}% Include figure files
\usepackage{dcolumn}% Align table columns on decimal point
\usepackage{bm}% bold math
\usepackage{wasysym}
\usepackage{textcomp}
\usepackage{epstopdf}
\usepackage{pxfonts}
\usepackage{txfonts}
\usepackage{tipa}
\usepackage{fontenc}
\usepackage{color,soul}
\usepackage{marvosym}
\begin{document}

\preprint{APS/123-QED}

\title{Crossover from paramagnetic compressed flux regime to diamagnetic pinned vortex lattice in a single crystal of cubic Ca$_3$Rh$_4$Sn$_{13}$}% Force line breaks with \\
\author{P. D. Kulkarni$^{1,2}$, S. S. Banerjee$^{3}$, C. V. Tomy$^{4}$, G. Balakrishnan$^{5}$, D. McK. Paul$^{5}$, S. Ramakrishnan$^{1}$ and A. K. Grover$^{1}$}
\affiliation{$^{1}$Department of Condensed Matter Physics and Materials
Science, Tata Institute of Fundamental Research, Homi Bhabha Road,
Colaba, Mumbai 400005, India.}
\affiliation{$^{2}$Laboratorio de Bajas Temperaturas, Departamento de Fisica de la Materia Condensada, Instituto de Ciencia de Materiales Nicolas Cabrera, Universidad Autonoma de Madrid, 28049, Madrid, Spain}
\affiliation{$^{3}$Department of Physics, Indian Institute of Technology Kanpur, Kanpur 208016, India}
\affiliation{$^{4}$Department of Physics, Indian Institute of Technology Bombay, Mumbai 400076, India}
\affiliation{$^{5}$Department of Physics, University of Warwick, Coventry CV4 7AL, United Kingdom}

\date{\today}% It is always \today, today,
\begin{abstract}
We report the observation of positive magnetization on field cooling (PMFC) in low  applied magnetic fields ($H < 100$\,Oe) in a single crystal of Ca$_3$Rh$_4$Sn$_{13}$ near its superconducting transition temperature ($T_c \approx 8.35$\,K). For 30\,Oe~$< H < 100$\,Oe, the PMFC response crosses over to a diamagnetic response as the temperature is lowered below 8\,K. For 100\,Oe~$< H < 300$\,Oe, the diamagnetic response   undergoes an unexpected reversal in its field dependence above a characteristic temperature (designated as $T^*_{VL} = 7.9$\,K), where the field-cooled cool-down magnetization curves  intersect. The in-phase and out-of-phase ac susceptibility data confirm the change in the superconducting state across $T^*_{VL}$. We ascribe the PMFC response to a compression of magnetic flux caused by the nucleation of superconductivity at the surface of the sample. In very low fields ($H< 20$\,Oe), the PMFC response has an interesting oscillatory behaviour which persists up to about 7\,K. The oscillatory nature underlines the interplay between competing responses contributing to the magnetization signal in PMFC regime. We believe that the (i) counterintuitive field dependence of the diamagnetic response for $H> 100$\,Oe and above $T^*_{VL}$ (lasting up to $T_c$), (ii) the oscillatory character in PMFC response at low fields and (iii) the PMFC peaks near 8.2\,K in 30\,Oe~$\leq H \leq 100$\,Oe provide support in favour of a theoretical scenario based on the Ginzburg-Landau equations. The scenario predicts the possibility of complex magnetic fluctuations associated with transformation between different metastable giant vortex states prior to transforming into the conventional vortex state as the sample is cooled below $T^*_{VL}$. 
\end{abstract}

\pacs{74.25.Ld, 74.25.Ha, 74.25.Op}% PACS, the Physics and Astronomy

\keywords{rare earth intermetallics, magnetic compensation, phase transition}

\maketitle
\section{Introduction}
\label{sec:INTRO}
Superconducting specimens of different genre and with varying pinning have been known \cite{Khomskii,Braunisch,Svedlindh,Reidling,Sigrist,Li,Thompson,Pust,Geim,Das} to display an anomalous paramagnetic response, instead of the usual diamagnetic Meissner effect, on field-cooling in  small magnetic fields ($H$). Such a response has been designated as the Paramagnetic Meissner Effect (PME) or the Wohlleben effect \cite{Khomskii}, since the advent of  superconductivity in cuprates. Originally this feature was found in granular \cite{Braunisch, Svedlindh} form of the high $T_c$ superconductor (HTSC) Bi$_2$Sr$_2$CaCu$_2$O$_8$ and in single crystals \cite{Reidling} of 
YBa$_2$Cu$_3$O$_7$. Invoking the possible special $d$-wave symmetry of the superconducting order parameter in HTSC materials, different models, such as, the presence of an odd number of $\pi$-junctions in a loop leading to spontaneous circulating currents producing a positive magnetization signal, the presence of Josephson junctions ($\pi$- or 0-), spontaneous supercurrents due to vortex fluctuations or an orbital glass \cite{Sigrist, Li}, were proposed to explain the PME. However, the subsequent observation of positive magnetization even in conventional $s$-wave superconductors, like, moderately pinned Nb discs \cite{Thompson, Pust}, nanostructured Al discs \cite{Geim} and a weakly pinned spherical single crystal \cite{Das} of Nb, has indicated that the origin of positive magnetization on field cooling in these materials is perhaps related to flux trapping \cite{Koshelev, Khalil, Moshchalkov, Zharkov} and its possible subsequent compression \cite{Koshelev, Moshchalkov, Zharkov}.  

Magnetic flux can get trapped in the bulk of a superconductor below $T_c$, as the preferential flux expulsion from the superconducting boundaries can lead to a flux free region near the sample edges, which would grow as the sample is further cooled \cite{Li,Das,Koshelev,Khalil,Moshchalkov,Zharkov,Saint}. In such a situation the magnetization response is governed by two counter flowing currents \cite{Li, Koshelev}, a paramagnetic (pinning) current flowing in the interior of the sample, which is associated with the pinned compressed flux, and a diamagnetic shielding current flowing around the surface of the sample, which screens the flux free region near the sample surface from the externally applied fields. Since these currents flow in opposite directions, the resultant magnetization can either be positive or negative \cite{Li, Koshelev, Zharkov}. 

Attempts to understand PME via the Ginzburg-Landau (GL) equations have shown \cite{Moshchalkov, Zharkov} that a large compression of magnetic flux in the interior of the superconductor is energetically equivalent to the creation of giant vortex states, with multiple  flux quanta $L\phi_0$, where the orbital quantum number, $L > 1$. Boundary effects in finite sized samples \cite{Moshchalkov, Zharkov, Fink} show that the Meissner state ($L = 0$ state) need not be the lowest energy state, but, a giant vortex state with $L > 0$ (in fact with $L > 1$) would have lower energy. Giant vortices are thus trapped inside the superconductor \cite{Moshchalkov} below a temperature where surface superconductivity \cite{Saint} is nucleated. Pinning may lead to a metastable giant vortex state with constant $L$ ($>1$) getting sustained without decay into $L$ states with lower energy\cite{Moshchalkov}, as the temperature is gradually reduced. On approaching the bulk superconductivity regime,  it is proposed theoretically \cite{Zharkov} that the transformation of a metastable giant vortex state into different lower $L$ states can lead to a magnetization response having the tendency to fluctuate between diamagnetic and paramagnetic values. 

In an earlier work \cite{Das}, some of the present authors reported the observation of surface superconductivity \cite{Saint} concurrent with positive magnetization on field cooling (PMFC) (often designated as Paramagnetic Meissner Effect (see Ref.~6)) in a weak pinning spherical single crystal ($r_0 \approx 1.1$\,mm) of Nb. However, there were no features in these experiments which could be ascribed to the metastable nature of giant vortex states in the temperature interval of the PMFC regime. In recent years, we have studied the ubiquitous Peak Effect (PE) phenomenon \cite{Pippard} in single crystals of a large variety of low $T_c$ and other novel superconductors \cite{Banerjee, Sarkar, Jaiswal,Tomy, Tomy1}. Amongst these,  the cubic stannide, Ca$_3$Rh$_4$Sn$_{13}$ ($T_c \sim 8.35$\,K)\cite{Tomy}, has a $\kappa \sim 18$. For this compound, we now present new and interesting results pertaining to the PMFC, emanating from the dc and ac magnetization measurements performed at low fields in close vicinity of $T_c$. The peak value of the paramagnetic signal in the field-cooled cool down (FCC) magnetization curves ($M_{FCC}(T)$) is inversely proportional to the magnetic field (10\,Oe~$ < H < 100$\,Oe) in which the sample is field-cooled. The paramagnetic signal close to $T_c$ at very low fields ($ H < 20$\,Oe) has a characteristic structure presenting a fluctuating response arising from competition between the paramagnetic and diamagnetic contributions. The ac susceptibility data also display interesting features, which appear consistent with the observations in dc magnetization measurements. A host of novel experimental findings reported here vividly illustrate the crossover from the compressed flux regime to the pinned conventional vortex lattice state, predicted and well documented by theorists in the literature \cite{Li, Moshchalkov, Zharkov}. 

\section{Experimental Details}
\label{sec:EXP}
The single crystals of Ca$_3$Rh$_4$Sn$_{13}$ were grown by the tin flux method \cite{Tomy}. Each growth cycle yielded a number of single crystals whose detailed pinning characteristics varied somewhat. The dc magnetization measurements were performed using a commercial SQUID-Vibrating Sample Magnetometer (Quantum Design (QD) Inc., USA, model S-VSM). In S-VSM, the sample executes a small vibration around a mean position, where the magnetic field is uniform and maximum. This avoids the possibility of the sample moving in an inhomogeneous field during the dc magnetization measurements. The remnant field of the superconducting magnet of S-VSM was estimated at different stages of the experiment, using a standard paramagnetic Palladium specimen. To ascertain the set value of the current supply  energising the superconducting coil to yield nominal zero field at the sample position, we also relied on the identification of the change in sign of the $z$-component of the magnetic field on its gradual increase (1 to 2\,Oe at a time) via independently examining the change in sign of the (field-cooled) magnetization values of the superconducting Sn specimen. The zero-field current-setting could thus be located to within $\pm1$\,Oe in a given cycle of gradual change (increase or decrease)  of field values from a given remnant state (positive or negative) of the superconducting magnet. The isofield temperature dependent magnetization curves were recorded by ramping the
temperatures in the range of 0.1\,K/min to 0.5\,K/min. The ac susceptibility measurements were carried out using another SQUID magnetometer (Q.D. Inc., USA, Model MPMS-5). The ac measurements were made at a frequency of 211\,Hz and  ac amplitude of 2.5\,Oe (r.m.s.). The applied fields in dc and ac measurements were kept normal to the plane of the rectangular platelet (1\,mm\,$\times$\,2\,mm\,$\times$\,1.5\,mm) shaped sample used in the present study. 

\section{Results and Discussion}
\label{sec:RES}
\subsection{Peak effect characteristic in magnetic hysterisis ($M$--$H$) isotherms}

\begin{figure}[ht]
\includegraphics[width=0.45\textwidth]{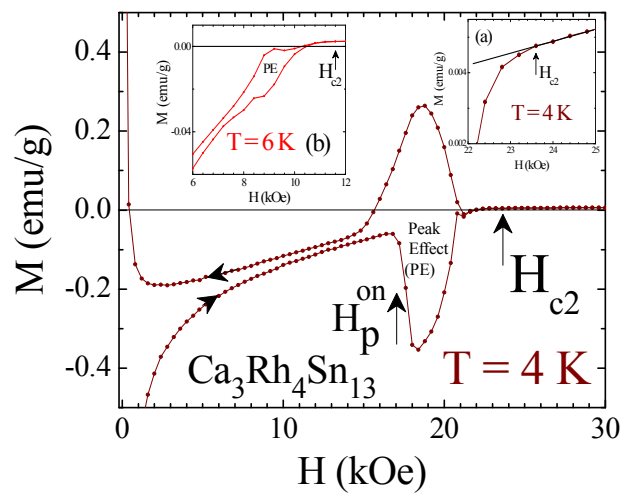}
\caption{\label{fig1}(colour online)  A portion of the dc magnetization hysteresis ($M$--$H$) curve at $T = 4$\,K in a single crystal of Ca$_3$Rh$_4$Sn$_{13}$. The upper critical field ($H_{c2}$) and the onset field of the PE ($H_p^{on}$) are marked. Arrows on the curve indicate the directions of the field change. The inset (a) shows the expanded portion of $M$--$H$ loop at 4\,K to identify the onset of superconductivity at $H_{c2}$. The inset (b) shows an expanded portion of the $M$--$H$ loop encompassing the PE region at 6\,K. }
\end{figure}

The main panel of Fig.~1 shows a portion of the isothermal $M$ vs $H$ loop recorded at $T = 4$\,K for a single crystal of Ca$_3$Rh$_4$Sn$_{13}$. The upper critical field ($H_{c2}$) and the onset field of the PE ($H_p^{on}$) are marked in the main panel. An anomalous enhancement of the magnetization hysteresis below $H_{c2}$ is a fingerprint of the peak effect (PE) phenomenon in Ca$_3$Rh$_4$Sn$_{13}$ \cite{Sarkar}.   The inset (a) in Fig.~1 elucidates the  deviation from linearity nucleating at  the
paramagnetic-superconductor boundary, taken as  $H_{c2}$. The inset (b) in Fig.~1 shows the PE region in a portion of the $M$ vs $H$ loop at 6\,K, with $H_{c2}$ marked as well. The second magnetization peak feature \cite{Sarkar} was not observed in the present sample. These data comprising only the PE attest to the high quality of the crystal \cite{Sarkar, Tomy} chosen for our present study. 

\subsection{Positive magnetization close to the onset of superconductivity in isofield scans at low fields}

An inset in Fig.~2 displays one of the typical temperature dependence of the $M_{FCC}(T)$ curves in low fields (viz., $H$ = 30\,Oe, here). $M_{FCC}$ signal can be seen to saturate to its diamagnetic limit at low temperatures ($T < 6$\,K). At the onset of the superconducting transition ($T_c = 8.35$\,K), $M_{FCC}(T)$ response is, in fact, paramagnetic, which is evident in the plots of the expanded $M_{FCC}(T)$ curves for $H = 30$\,Oe, 60\,Oe and 90\,Oe (see main panel of Fig.~2). The paramagnetic magnetization on field cooling (PMFC) in a given field ($H \leq 100$\,Oe) reaches a peak value before turning around to crossover towards diamagnetic values (near 8\,K). The PMFC data for 30\,Oe~$\leq$ $H$ $\leq 90$\,Oe in Fig.~2 reveal that (i) the height of the paramagnetic peak decreases monotonically as $H$ increases and (ii) the competition between positive signal and the diamagnetic shielding response gives rise to the turnaround behaviour in PMFC signals near 8.15\,K. No significant difference was noted between PMFC response at $T > 8.2$\,K for $H < 100$\,Oe in the data recorded (not shown here) during the field-cooled warm-up (FCW) and FCC modes. This in turn implies that the positive magnetization signals above about 8.2\,K do not depend on the  thermomagnetic history of the applied magnetic field. This led us to explore closely the isothermal magnetization hysteresis loops  in the temperature range 8\,K~$< T < 8.35$\,K. 

\begin{figure}[ht]
\includegraphics[width=0.45\textwidth]{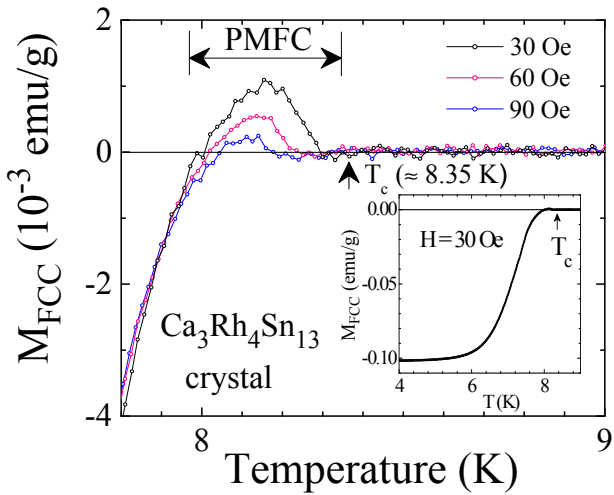}
\caption{\label{fig2}(colour online) The inset panel shows the temperature variation of $M_{FCC}$ in $H = 30$\,Oe in a single crystal of Ca$_3$Rh$_4$Sn$_{13}$. The main panel shows portions of $M_{FCC}(T)$ curves at $H$ =  30\,Oe, 60\,Oe and 90\,Oe.}
\end{figure}

\begin{figure}[ht]
\includegraphics[width=0.45\textwidth]{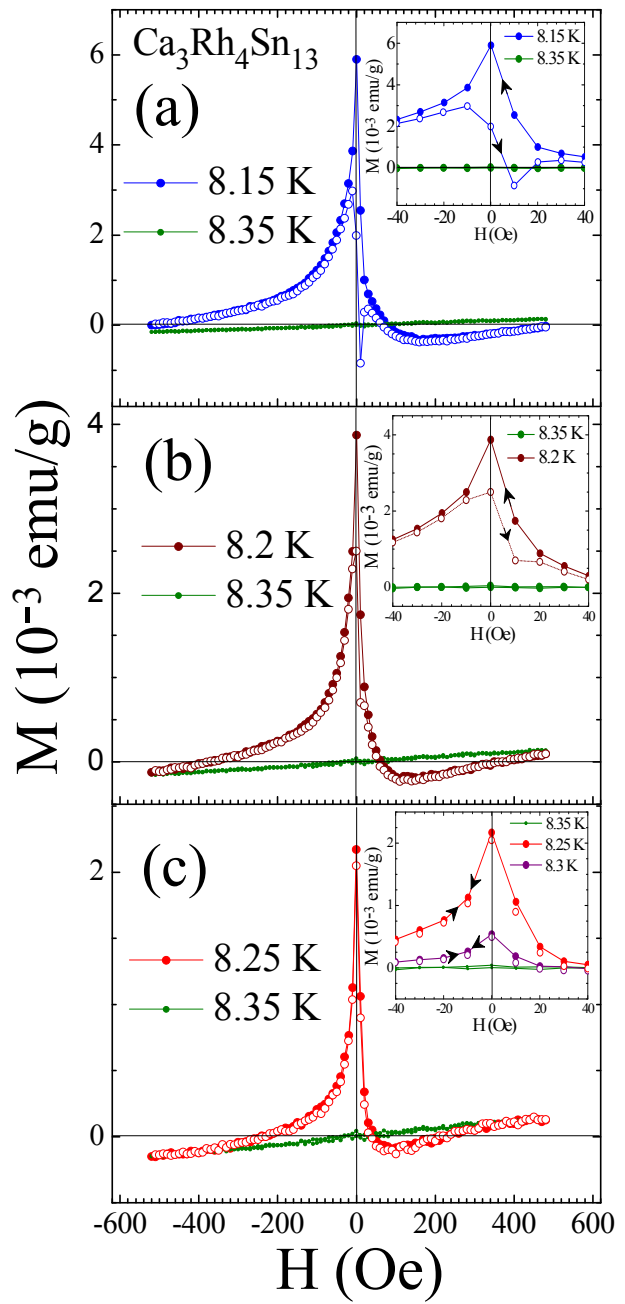}
\caption{\label{fig3}(colour online) Magnetization vs applied magnetic field at (a) 8.15\,K, (b) 8.2\,K and (c) 8.25\,K, in Ca$_3$Rh$_4$Sn$_{13}$. In each of the panels,  $M$ vs $H$ plot at 8.35\,K is also shown. The filled/open circles  in each panel represent scans from +/$-$500\,Oe to $-$/+500\,Oe, respectively. Inset panels show the $M$--$H$ data across the zero field region on expanded scales. }
\end{figure}

Figures 3(a) to 3(c) show the pair-wise plots of $M$--$H$ data recorded at 8.15\,K, 8.20\,K and 8.25\,K in comparison with the corresponding data at 8.35\,K. For each curve, the sample was cooled down to a given temperature in a field of +500 Oe. The field was then repeatedly ramped between $\pm 500$\,Oe. Note first that the $M$--$H$ data at 8.35\,K is linear and passes through the origin, as anticipated, since this temperature identifies the onset of the superconducting transition ($T_c = 8.35$\,K). The pair of $M$--$H$ plots (at 8.25\,K and 8.35\,K) in Fig.~3(c) reveal that even at 8.25\,K, a diamagnetic response (as determined by the difference between the two plots) is clearly present at about 250\,Oe. On lowering the field below about 40\,Oe, a sharp upturn takes the magnetization from diamagnetic to paramagnetic values. The paramagnetic response reaches its peak value at the zero applied field (in the $z$-direction). The peak value of the paramagnetic signal in zero field is seen to decrease with enhancement in field on either side of the zero field. An inset in Fig.~3(c) shows a comparison of the field variation of the paramagnetic response at 8.25\,K and 8.30\,K on either side of the zero field on an expanded scale. Note the asymmetry in the field variation of the paramagnetic response at positive and negative fields. The observed asymmetry at 8.25\,K and 8.3\,K is independent of whether the sample is cooled first in +500\,Oe or $-$500\,Oe. We believe that the paramagnetic response at zero field (in $z$-direction), which is superconducting in origin, reflects the magnetization signal due to compression of field corresponding to $x$- and $y$- components of the earth's field. The  magnetization value at zero field (in $M$--$H$ loops) is found to be larger at 8.25\,K as compared to that at 8.3\,K (cf. inset in Fig.~3(c)). Such an enhancement  characteristic can be seen to continue at a further lower temperature of  8.2\,K (see inset panel of Fig.~3(b)). An inset panel in Fig.~3(a) shows the $M$--$H$ plot at 8.15\,K on the expanded scale across the zero-field region. From this inset panel, it is apparent that the $M$--$H$ loop at 8.15\,K has started to imbibe the characteristic of a hysteretic magnetic response in the neighbourhood of nominal zero field. The $M$--$H$ loop at 8.15\,K in the inset of Fig.~3(a), therefore, appears to be a superposition of (i) a hysteretic $M$--$H$ loop expected in a type-II superconductor and (ii) a PMFC signal decreasing with enhancement in field on either side of the nominal zero-field.
 
The PMFC signal in $M(T)$ measurements in Fig. 2 for Ca$_3$Rh$_4$Sn$_{13}$ is an important observation at $H < 100$\,Oe and $T > 8$\,K. Above 100\,Oe, the magnetization response in the superconducting state (at $T < 8.35$\,K) is largely diamagnetic, however, an important unexpected change is witnessed in the field dependence of the diamagnetic response in the neighbourhood of 8\,K, as described ahead.

\subsection{Crossover from compressed flux regime to pinned vortex lattice regime below 8\,K}

\begin{figure}[ht]
\includegraphics[width=0.45\textwidth]{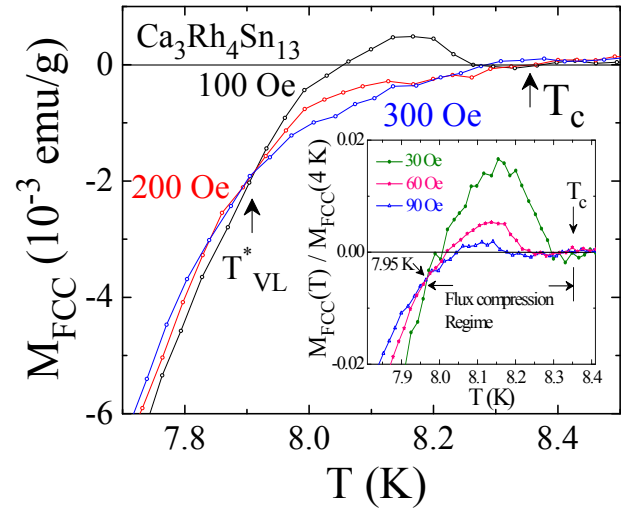}
\caption{\label{fig4}(colour online) Portions of the field-cooled cool down magnetization as a function of temperature at $H$ = 100\,Oe, 200\,Oe and 300\,Oe in Ca$_3$Rh$_4$Sn$_{13}$. The transition temperature $T_c$ and the temperature corresponding to the intersection of the three $M_{FCC}(T)$ curves (identified as $T^*_{VL}$) are marked in the main panel. The inset panel shows the temperature dependences of the normalized magnetization (see text) at lower fields, $H < 100$\,Oe. The flux compression region has been identified in the temperature interval from $T_c$ down to the crossover temperature.}
\end{figure}

The main panel in Fig.~4 displays the portions of the $M_{FCC}(T)$ curves close to $T_c$ in $H$ = 100\,Oe, 200\,Oe and 300\,Oe in Ca$_3$Rh$_4$Sn$_{13}$. The most striking feature of these data is the intersection of the $M_{FCC}(T)$ curves at 7.9\,K (identified as $T^*_{VL}$). Below $T^*_{VL}$, the magnitude of the diamagnetic response decreases as the field increases, as expected for the vortex lattice (VL) in a type-II superconductor. However, for 7.9\,K\,$< T < 8.35$\,K, the magnitude of the diamagnetic response is enhanced as the field increases, which is unusual for a conventional low-$T_c$ type-II superconductor. Such a behaviour, however, has been reported \cite{Grover, Kadowaki, Kes} in the context of a high-$T_c$ Josephson-coupled layered superconductor (JCLS) Bi$_2$Sr$_2$CaCu$_2$O$_{8-\delta}$ (Bi2212) for $H \parallel  c$,  where a crossover happens at a corresponding $T^*$ value between the type-II response of a JCLS  and the superconducting fluctuations-dominated response of the decoupled pancake vortices. In the case of Ca$_3$Rh$_4$Sn$_{13}$, the crossover at $T^*_{VL}$ is, however, between the pinned vortex lattice state (VL) and the compressed flux regime, giving rise to PMFC signals at $H < 150$\,Oe in the neighbourhood of $T_c$. We identify the region between $T_c$ and $T^*_{VL}$ as the flux compression region.

The $M_{FCC}(T)$ curves for $H < 100$\,Oe, shown in Fig.~2, did not intersect at a unique temperature. However, if the $M_{FCC}(T)$ curves (for 30\,Oe~$\leq H \leq 90$\,Oe) are normalized  to their respective values at 4\,K, we observe a crossover at 7.95\,K (see the inset panel of Fig.~4). Below 7.95\,K, the response of the normalized $M_{FCC}(T)$ curves for different $H$ is like that in a  pinned type-II superconductor, and above 7.95\,K, there exists the compressed flux regime\cite {Koshelev,Moshchalkov}, accounting for the positive peaks in magnetization above 8\,K and up to $T_c$.

\begin{figure}[ht]
\includegraphics[width=0.45\textwidth]{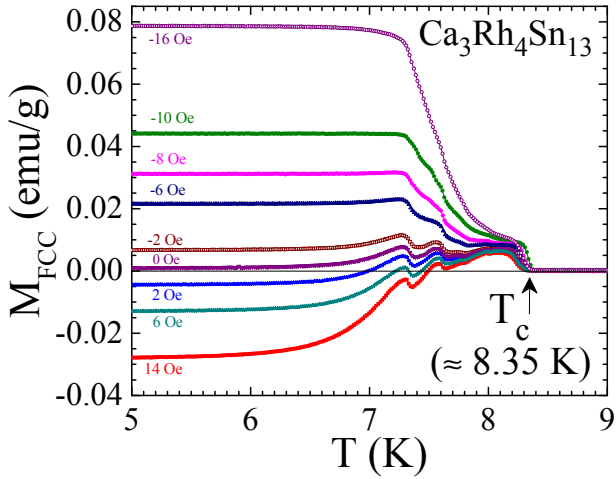}
\caption{\label{fig5}(colour online) Temperature dependences of the field-cooled cool down magnetization (M$_{FCC}(T)$) values measured in Ca$_3$Rh$_4$Sn$_{13}$ by progressively incrementing the applied
field from $-$16 Oe to +14 Oe.}
\end{figure}

\subsection{Oscillatory behaviour in field-cooled magnetization curves at low fields}

Figure 5 summarizes the $M_{FCC}$ data sequentially recorded from $H$ = $-$16\,Oe to +14\,Oe in the single crystal of Ca$_3$Rh$_4$Sn$_{13}$.
The sample was initially cooled in the remnant field of the superconducting magnet, whose value was estimated by measuring the paramagnetic magnetization of the standard Pd sample. The current in the superconducting coil was then incremented step-wise so as to enhance magnetic field by 2\,Oe each time. The following characteristics are noteworthy in Fig.~5:
(i) While in positive fields ($H \geq 2$\,Oe), the PME signal close to $T_c$ gives way to diamagnetic Meissner response at lower temperatures, in negative fields, the same PME signal close to $T_c$ adds on to  the  positive Meissner response at lower temperatures. Thus, there is no change in the sign of magnetization response as a function of temperature in negative fields. The anomalous PME peak feature prominently evident at positive fields, therefore, takes the form of an onset of a sharper upturn in magnetization below 8\,K in negative fields.
(ii) A vivid oscillatory character is present below 8.2\,K in the $M_{FCC}(T)$ curves for fields ranging from $H$ = $-$6\,Oe to +14\,Oe. On the negative field side, the oscillatory feature tends to get obscured at $H = -10$\,Oe. Details of the oscillatory structure between 8.2\,K and 7\,K depend somewhat on the rate of cooling down while recording the $M_{FCC}$ data in a given field. 

The asymmetry in response in the isofield runs in positive and negative fields ($| H | < 30$\,Oe) in Fig.~5 correlates with the asymmetry in the response evident in $M$--$H$ isotherms shown in the inset panels of Fig.~3. The fact that the positive signal at nominal zero fields  decays with field on either side of the zero-field (cf. plots at 8.25\,K and 8.30\,K in the inset of Fig.~3(c)) implies that the signal would not change sign in $M_{FCC}(T)$ curves measured for  negative applied magnetic fields. For negative magnetic field, the diamagnetic shielding response emanating from a usual pinned type-II superconducting state would result in a  positive signal in the magnetization measurements. Such a positive signal superposed on the PMFC magnetization signal (decaying with field) would rationalise the absence in the change of the sign of the PME signal in negative fields in temperature dependent scans.  

\begin{figure}[ht]

\includegraphics[width=0.45\textwidth]{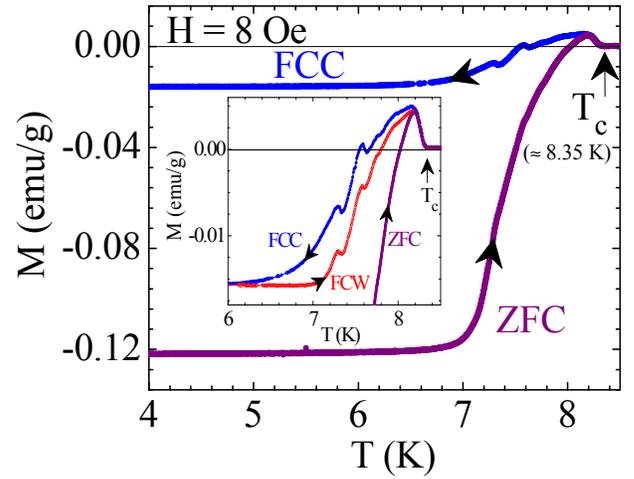}
\caption{\label{fig6}(colour online) Temperature variation of the zero field-cooled (ZFC) and field-cooled (FC) dc magnetization curves in $H$ = +8\,Oe  in Ca$_3$Rh$_4$Sn$_{13}$. The inset shows $M_{ZFC}$, $M_{FCC}$ and $M_{FCW}$
(field-cooled warm up)  plots in $H$ = +8\,Oe on an expanded scale near $T_c$.}
\end{figure}

Figure 6 shows a comparison of zero field-cooled (ZFC) magnetization response, $M_{ZFC}(T)$, in $H = 8$\,Oe along with its $M_{FCC}(T)$ run. To record the $M_{ZFC}(T)$ run, the Ca$_3$Rh$_4$Sn$_{13}$ crystal was initially cooled down to 4\,K in (estimated) zero field, the field was then incremented by +8\,Oe and the magnetization was measured while  slowly increasing the temperature  above $T_c$. The crystal was then cooled down to 4\,K to record the 
$M_{FCC}(T)$ data, and thereafter the magnetization was  once again measured in the warm-up mode $M_{FCW}(T)$ to temperatures above $T_c$. The inset panel in Fig.~6 shows the plots of $M_{FCC}(T)$, $M_{FCW}(T)$ and $M_{ZFC}(T)$ close to $T_c$ on an expanded scale. It can be noted that while the oscillatory characteristic is evident in $M_{FCC}(T)$ and $M_{FCW}(T)$ runs, the $M_{ZFC}(T)$ is devoid of any oscillatory modulation feature as the diamagnetic response (in +8\,Oe) crosses over to yield the attribute of PME peak between 8\,K and 8.35\,K. The three curves in the inset panel of Fig.\,6 meet near 8.2\,K, above which the path independent paramagnetic response monotonically decreases. It is reasonable to state that during the ZFC run in $H = 8$\,Oe, the quantized vortices will enter the sample at a temperature at which the lower critical field $H_{c1}(T)$ becomes less than 8\,Oe 
(ignoring the surface barrier effects). The quantized vortices will distribute inside the sample to yield Bean's Critical State \cite{Bean} profile and the macroscopic currents $J_c(B)$ will flow inside the sample. The onset of sharp fall in $M_{ZFC}(T)$ above 7\,K reflects the decrease in $J_c(B)$ with $T$ on approaching the superconducting transition temperature. 

It is tempting to associate the oscillatory responses in Fig.~5 to the notion of competition between the (Abrikosov) quantized vortices splitting out of the giant vortex state(s) in the form of compressed flux, and the tendency of a given giant vortex to retain (i.e., conserve) its angular momentum \cite{Moshchalkov} due to pinning. The high $\kappa$ of Ca$_3$Rh$_4$Sn$_{13}$ ordains that the different $L$ states of the giant vortex are closely spaced in energy. By lowering the temperature, there is a tendency to transform from $L > 1$ state to $L = 1$ state (Abrikosov state). However, theoretical work \cite{Moshchalkov,Zharkov} has shown that, due to pinning, the system can exhibit metastability, wherein, there can be fluctuations in magnetization corresponding to the transformation between different metastable $L$ states before the system attains the $L = 1$ state. 

In the framework of GL equations yielding multi-flux quanta, the magnetization due to different $L$ states follow different temperature dependences at different reduced fields (i.e., applied field normalized to the thermodynamic
critical field, $H_c$). In very low reduced fields ($h$) (e.g., $h \approx 0.001$, $\kappa \approx 10$ and cylindrical geometry), it has been calculated \cite{Moshchalkov}  that all the $L$ states will make paramagnetic contributions
such that higher $L$ values contribute more. In the case of Ca$_3$Rh$_4$Sn$_{13}$, where $H_c \approx 3$\,kOe \cite{Sarkar}, the PMFC response is observed in the range of reduced fields,  10$^{-3}$ to  10$^{-2}$, where 
contributions from $L \geq 1$ states slightly below $T_c$ could be paramagnetic. If the possible transitions between different high $L$ states occur at the same temperature in the very low $h$ range, one could rationalize the insensitivity
of oscillatory pattern to the applied fields in Fig.~5. We may also add here that in the GL scenario, the irreversibility temperature is argued \cite{Moshchalkov} to correspond to a crossover between giant vortex states and the Abrikosov quantized vortices, consistent with the observations shown in Fig.~6.

The difference in the (diamagnetic) magnetization behaviour in FCC and FCW modes had been noted in samples of conventional low-$T_c$ \cite{Tomy2} and high $T_c$ \cite{Deak} superconductors. Clem and Hao \cite{Clem} had shown how it could be rationalized in the framework of the Critical State Model \cite{Bean}. The spatial distribution of macroscopic currents ($J_c(B)$, where $B$ is the local magnetic field) that are set up within an irreversible type-II superconductor while cooling down is different from that which emerges while warming up the sample in the same external field. The diamagnetic $M_{FCW}$ curve typically lies below the $M_{FCC}$ curve, and the two curves merge at the irreversibility temperature \cite{Clem}, where $J_c(B)$ vanishes. In high $T_c$ superconductors the irreversibility line lies well below the $H_{c2}$ line. In strongly pinned samples of type-II superconductors, the irreversibility temperature $T_{irr}(H)$ approaches 
$T_c(H)$ \cite{Grover1}. In this context, the merger of $M_{FCW}$ and $M_{FCC}$ curves in $H = 8$\,Oe (cf. inset, Fig.~6) could imply that the macroscopic $J_c$ ($B = 8$\,Oe) approaches zero just above 8.2\,K. We may further add that the overlap of $M_{FCC}$ and $M_{FCW}$ curves at $T > 8.2$\,K in Fig.~6 and the behaviour of $M$ vs $H$ at 8.25\,K in Fig.~3(c) validates the theoretical prediction \cite{Moshchalkov} that the PMFC signal first decays rapidly with field, followed by the emergence of a diamagnetic response at higher fields. 

\subsection{AC susceptibility measurements in Ca$_3$Rh$_4$Sn$_{13}$}
\begin{figure}[ht]
\includegraphics[width=0.47\textwidth]{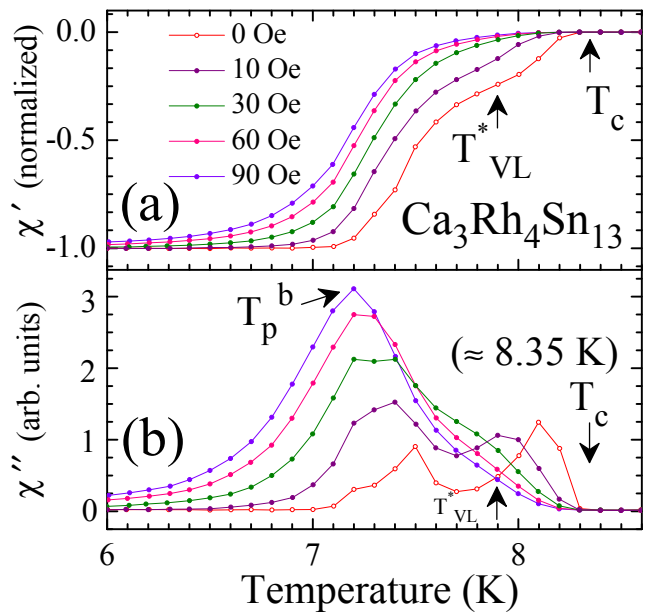}
\caption{\label{fig7}(colour online) Panels (a) and (b), respectively, show the temperature variation of the in-phase ($\chi^\prime$) and out-of-phase ($\chi^{\prime\prime}$) ac susceptibility in Ca$_3$Rh$_4$Sn$_{13}$ at different fields, as indicated.}
\end{figure}

Figures 7 and 8 summarize the in-phase ($\chi^\prime$) and out-of-phase ($\chi^{\prime\prime}$) ac susceptibility data recorded in h$_{ac}$ of 2.5 Oe (r.m.s.) iso-field and iso-thermal runs, respectively. The iso-field runs were made while cooling down from the normal state ($T > 8.35$ K). The isothermal data were recorded along four or five quadrants within the field limts of $\pm$ 200 Oe, for the sample having been initially cooled in nominal zero field or + 500 Oe, respectively.

Figures 7(a) and 7(b) show the $\chi^{\prime}(T)$ and $\chi^{\prime\prime}(T)$ plots recorded while cooling down the Ca$_3$Rh$_4$Sn$_{13}$ crystal in dc fields of  0\,Oe (nominal value), 10\,Oe, 30\,Oe, 60\,Oe and 90\,Oe, respectively. The $T_c$ and $T^*_{VL}$ values stand marked appropriately in these two panels. The $\chi^{\prime}$ response below as well as above $T^*_{VL}$ remains diamagnetic. However, a conspicuous change in temperature dependence of $\chi^{\prime}$ can be noted to happen near $T^*_{VL}$. Such a change is often ascribed \cite{Das} to the crossover between the shielding response in the bulk to the shielding response from the surface superconductivity. In the present case, where we witness the PMFC signal above 8\,K in dc magnetization data, it can be noted that $\Delta M/\Delta H$ is negative (cf. Fig. 4), which rationalizes the diamagnetic $\chi^{\prime}$ response above $T^*_{VL}$ in Fig.~7(a).  

\begin{figure}[ht]
\includegraphics[width=0.47\textwidth]{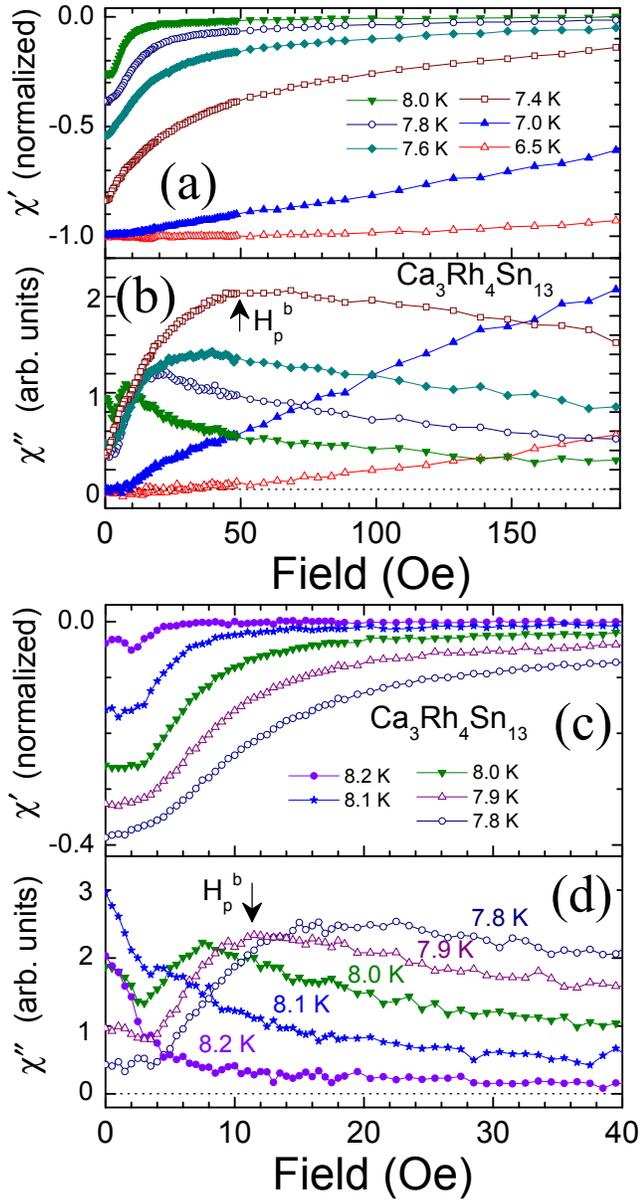}
\caption{\label{fig8}(colour online) Panels (a) and (b), respectively, show the field variation of the in-phase ($\chi^{\prime}$) and out-of-phase ($\chi^{\prime\prime}$) ac susceptibility in Ca$_3$Rh$_4$Sn$_{13}$ at the temperatures as indicated. Panels (c) and (d), respectively, show the field dependence of $\chi^{\prime}$ and $\chi^{\prime\prime}$ at selected temperatures on going across $T^*_{VL}$.}
\end{figure}
 
\begin{figure}[ht]
\includegraphics[width=0.47\textwidth]{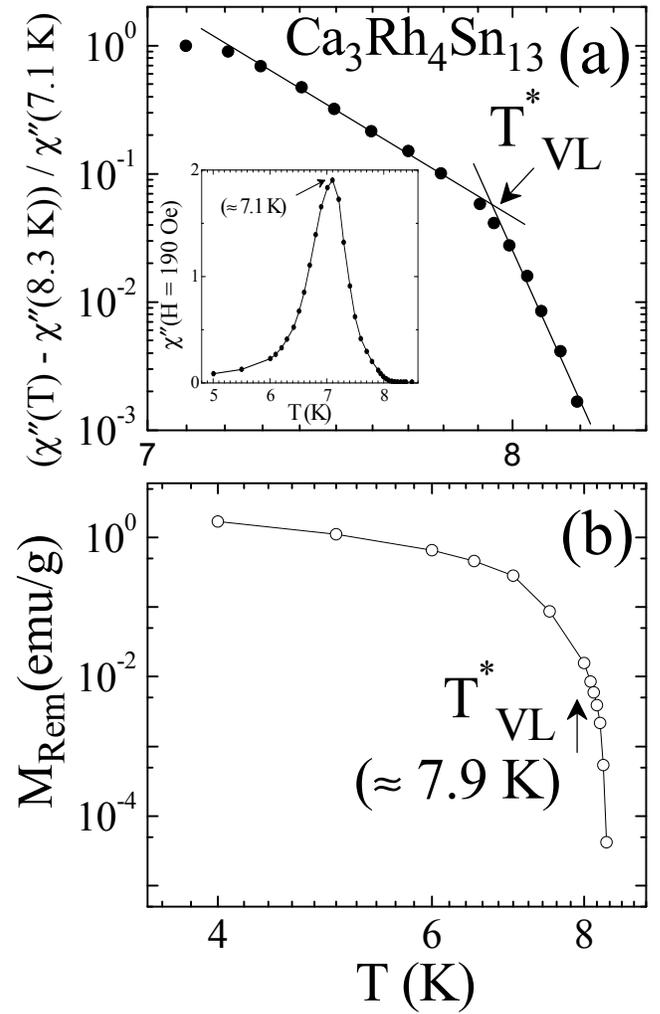}
\caption{\label{fig9} Inset in panel (a) shows  $\chi^{\prime\prime}$ (at $H = 190$\,Oe) as a function of temperature in Ca$_3$Rh$_4$Sn$_{13}$ showing the peak temperarure at  7.1\,K. Panel (a) shows temperature variation of the normalized $\chi^{\prime\prime}$ (see text) and  panel (b) shows   $M_{rem}$ values  estimated from the dc magnetization loops like those given in Fig.~3.}
\end{figure}

The $\chi^{\prime\prime}(T)$ data in Fig.~7(b) shows a dissipation response measured with an ac amplitude of 2.5\,Oe (r.m.s.) on either side of $T^*_{VL}$ of 7.9\,K. The two peaks of $\chi^{\prime\prime}(T)$ curve in nominal zero dc field in Fig.~7(b) support the notion of a crossover from superconductivity in the bulk (below 7.9~K) to the compressed flux regime (above it). The peak intensity of the higher temperature peak (above 7.9\,K) diminishes as the field increases from 10\,Oe to 60\,Oe. This correlates with the decline in the paramagnetic response with enhancement in field in the temperature regime of compressed magnetic flux (cf. Fig.~2 and Fig.~3(c)). A comparison of $\chi^{\prime\prime}(T)$ curves from $H = 0$\,Oe to 90\,Oe below 7.9\,K reveals that the lower temperature dissipative peak progressively becomes more prominent and the peak temperature moves inwards with the enhancement in dc field. This is the usual behaviour expected for enhanced irreversibility on cooling due to macroscopic currents set up within the bulk of a pinned type-II superconductor. The field ($H$) dependence of the peak temperature ($T_p^b$) of the dissipative peak below 7.9\,K can easily be rationalized in terms of field/temperature dependence of macroscopic currents ($J_c(B,T))$, flowing as per Critical State Model\cite{Bean} in the bulk of the sample.  
 
\begin{figure}[ht]
\includegraphics[width=0.47\textwidth]{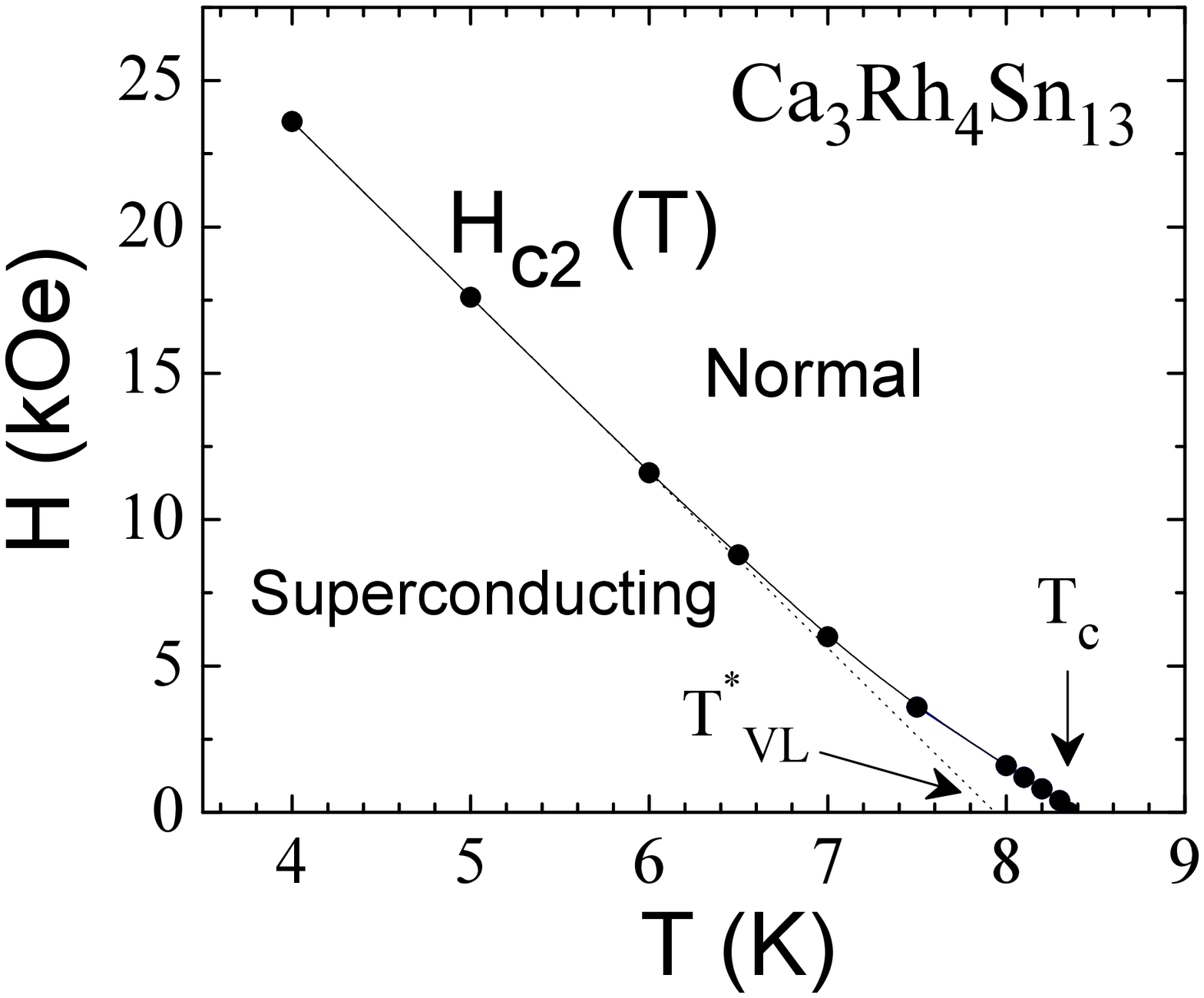}
\caption{\label{fig10} The $(H,T)$ phase diagram in a single crystal of Ca$_3$Rh$_4$Sn$_{13}$. The dashed line represents linear extrapolation of $H_{c2}(T)$ data. The temperature $T^*_{VL}$ is marked at 7.9\,K using the data in
 Fig.~2.}
\end{figure}

Hysteretic behaviour in isothermal $\chi^{\prime}$($H$) and $\chi^{\prime\prime}$($H$) were present at the $T$ \textgreater~7.5 K, however, the qualitative feature in field dependence of $\chi^{\prime}$($H$) and $\chi^{\prime\prime}$($H$) during field ramp-up or ramp-down remained the same. Above 8 K, $\chi^{\prime}$($H$) and $\chi^{\prime\prime}$($H$) data did not display significant hysteresis. To facilitate the comparison with the dc magnetization data in Fig. 3, we show in Figs. 8(a) to 8(d) the $\chi^{\prime}$ vs $H$ and $\chi^{\prime\prime}$ vs $H$ data recorded at selected temperatures below and above $T^*_{VL}$ of 7.9\,K for field ramp down from +200 Oe to 0 Oe, for sample having cooled in +500 Oe. The $\chi^{\prime}(H)$ response at 6.5\,K in Fig.~8(a) shows that the given $h_{ac}$ is almost completely shielded up to a dc field of 200~Oe. On raising the temperature to 7.0~K, the decline in $|~\chi^{\prime}|$ vs $H$ in Fig.~8(a) reflects the field dependence of $J_c(B)$ at that temperature. The same trend continues on raising the temperature upto about 8.0\,K. The $\chi^{\prime\prime}$ vs $H$ response at $T = 6.5$\,K in Fig.~8(b) confirms that the $h_{ac}$ of 2.5\,Oe is not able to yield appreciable dissipation inside the sample upto a dc field of 200\,Oe. However,  $\chi^{\prime\prime}$ vs $H$ response at 7.0\,K clearly reveals the presence of dissipative peak at a dc field of about 50\,Oe (marked as $H_p^b$). Thereafter, the decrease in $\chi^{\prime\prime}$ vs $H$ reflects the field dependence of $J_c(B)$.

A very interesting behaviour in $\chi^{\prime\prime}$ vs $H$, however, emerges (see Fig.~8(d)) as the temperature is raised from 7.8\,K upto 8.1\,K and beyond. The $\chi^{\prime}$ vs $H$ response at $T \geq 8.0$\,K in Fig.~8(c) indicates that for the given $h_{ac}$, $\chi^{\prime}$ has somewhat feeble field dependence at very low dc field ($H < 5$\,Oe). The $\chi^{\prime\prime}$ vs $H$ curves in Fig.~8(d), however, reveal that a qualitative change in very low field ($H < 5$\,Oe) response occurs at temperatures above 7.9\,K. Note that $\chi^{\prime\prime}$ vs $H$ curves at 8.1\,K and 8.2\,K in Fig.~8(d) show the dissipation is maximum at nominal zero field, and it decreases rapidly on enhancing the dc field. The $\chi^{\prime\prime}$ vs $H$ curve at 8.0\,K in Fig.~8(d) can be seen to imbibe the feature of a rapid decline of dissipation (which is maximum at zero field) with field, followed by surfacing of the dissipation peak (at $H_p^b$) due to currents in the bulk of the sample. The data in Fig.~8(d), therefore, illustrate once again the crossover from a pinned type-II superconducting state to the compressed flux regime across the temperature region of about 8\,K. The enhanced dissipation near zero field above 8.1\,K perhaps indicates the  dissipation from giant vortex cores with large $L$ nucleated by surface superconductivity, whose evidence we have already shown in Fig.~7(b).   

The inset of Fig.~9(a) shows a plot of $\chi^{\prime\prime}$ vs $T$ measured with an $h_{ac}$ of 2.5\,Oe (r.m.s.) in a dc field of 190\,Oe. The observation of a peak in $\chi^{\prime\prime}(T)$ at 7.1\,K implies that the given $h_{ac}$ fully penetrates the bulk of the sample at this temperature in $H_{dc} = 190$\,Oe. The decrease in $\chi^{\prime\prime}(T)$ above 7.1\,K reflects the usual decrease in $J_c$ with an increase in $T$. One can use this information to compute a relative dissipative response at $H=190$\,Oe w.r.t. the dissipation at the same field close the normal state, i.e, at 8.3\,K [($\chi^{\prime\prime}(T)-\chi^{\prime\prime}(8.3$\,K)]/$\chi^{\prime\prime}$(7.1\,K). This, in turn, amounts to computing the relative values of $J_c$ in a field of 190\,Oe w.r.t. its value at 7.1\,K. The main panel of Fig.~9(a) shows a plot of the above stated relative response as a function of temperature. Note a change in the slope of the plotted curve at about 7.9\,K (the so called $T^*_{VL}$ value). We believe that the region beyond 7.9\,K identifies the temperature dependence of surface pinning.

We have also plotted the remnant magnetization (or peak magnetization in close vicinity of the nominal zero field) determined from the $M$--$H$ loops (as in Fig.~3) as a function of temperature in Fig.~9(b). Such a remnant value ($M_{rem}$) could be taken as indicative of overall pinning in the specimen. We have marked the location of $T^*_{VL}$ (= 7.9\,K) in the semi-log plot of $M_{rem}$ vs $T$ in Fig.~9(b) to focus attention onto setting in of more rapid decline in $M_{rem}(T)$ on going across from (irreversible) pinned vortex lattice to paramagnetic compressed flux regime, where the remnant signal provides a measure of the dominance of the paramagnetic current. {%In Fig.9, we have chosen to label the two straight lines intersecting at $T^*_{VL}$ with pinning in the bulk and compressed flux regime, respectively. We believe that the comparison of $M_{rem}$ in Fig. 9(b) above $T^*_{VL}$ with the change in curvature of the $J_c$($T$) behaviour in Fig. 9(a) indicates that surface effects play a significant role in generating the PME response.}  

\section{Summary and Conclusion}

We have presented the results of dc and ac magnetization measurements at low fields in a weakly pinned single crystal of a low $T_c$ superconductor, Ca$_3$Rh$_4$Sn$_{13}$, which crystallizes in a cubic structure. This system had been in focus earlier\cite{Sarkar} for the study of the order-disorder transformation in vortex matter (at $H > 3$\,kOe) via the peak effect phenomenon. New results at very low fields and in close proximity of $T_c$ have revealed the presence of positive dc magnetization on field cooling. In $H < 20$\,Oe, PMFC signals nucleating at 8.35\,K can be seen to survive down to about 7\,K. For 30\,Oe~$ < H < 100$\,Oe, the crossover from paramagnetic magnetization values to diamagnetic values is seen to occur near 8\,K. For 100\,Oe~$\leq H \leq 300$\,Oe, the field cooled magnetization curves are observed to intersect at a temperature of 7.9\,K, below which the diamagnetic response is akin to that expected for a pinned vortex lattice in a type-II superconductor. We have attributed the PMFC response to the notion of compressed flux trapped within the body of the superconductor. Below 20\,Oe, the surfacing of a curious oscillatory structure in the PMFC response prompted us to invoke the possible notion of a conservation of angular momentum for the giant vortex state \cite{Moshchalkov, Zharkov} to account for this behaviour. The iso-field and iso-thermal ac susceptibility ($\chi^{\prime}$ and $\chi^{\prime\prime}$) data also seem to register the occurrence of a crossover between the compressed flux regime and the pinned vortex lattice. 

To conclude, we show in Fig.~10 the plot of $H_{c2}$ values as a function of temperature in the form of a magnetic phase diagram in which the normal and superconducting regions are identified. Between 4\,K and 7\,K, $H_{c2}$ versus $T$ has a linear variation; on extrapolation, this linear behaviour fortuitously meets the $T$-axis (where $H = 0$) at $T^*_{VL}$ of 7.9\,K. For $H < 300$\,Oe, the fingerprints of a compressed flux regime in the form of PMFC and/or anomalous diamagnetic response ($\Delta M/\Delta H < 0$) can be observed between $T_c$ and $T^*_{VL}$ of 7.9\,K. The region between $H_{c2}(T)$ line and the dotted line which meets the temperature axis at $T^*_{VL}$ in Fig.~10 is the regime where we have identified the presence of surface superconductivity and surface pinning (cf. Fig.~9). If this were so, then the portion of $H_{c2}(T)$ which deviates from the extrapolated dotted line in Fig.~10, should be identified as a portion of the $H_{c3}(T)$ line. At somewhat below $T^*_{VL}$ (e.g., at $T = 7.7$\,K), an estimate of the ratio of fields associated with the dotted portion of the line and that of the $H_{c2}(T)$ line gives a value of about 2 which is more like the ratio of $H_{c3}(T)/H_{c2}(T)$. In a spherical single crystal of elemental Nb, whose $\kappa$ value ($\sim 2$) was just above the threshold for type-II response, some of us had reported\cite{Das} the observation of surface superconductivity concurrent with the PMFC response over a large ($H,T$) domain, such that the $H_{c3}(T)$ was distinctly different from $H_{c2}(T)$ line in its phase diagram (Fig.~4 in Ref.~10). In the present case of Ca$_3$Rh$_4$Sn$_{13}$, where $\kappa$ is large ($\sim 18$), the PMFC signal, presumably sustained by the nucleation of superconductivity at the surface, is present only at low fields and in the close proximity to $T_c$. A sharp distinction between $H_{c3}$ and $H_{c2}$ is not discernible near $T_c$, the surface superconductivity could, however, be responsible for the slight concave curvature of the $H_{c2}(T)$ curve near $T_c$ in the magnetic phase diagram (cf. Fig. 10). 

We believe that behaviour reported above in Ca$_3$Rh$_4$Sn$_{13}$ is generic. Similar features (In particular, an apparent absence of PME peak like feature in negative applied fields and the associated asymmetry between responses
in positive/negative fields)  would be present in other weak pinning superconductors. 
Preliminary searches in single crystal samples of other superconducting compounds, like, Yb$_3$Rh$_4$Sn$_{13}$, NbS$_2$, etc. have yielded positive indications \cite{Grover2}.

\noindent The single crystals grown at University of Warwick form a part of the continuing programme supported by EPSRC of U.K. We thank Mahesh Chandran for fruitful discussions. SSB would like to acknowledge the funding from Indo-Spain Joint Programme of co-operation in S \& T, DST, India.

\end{document}